\documentclass[sigconf, nonacm]{acmart}
\usepackage{graphicx}
\usepackage{latexsym}
\usepackage{listings}
\usepackage{xcolor}
\usepackage{amsthm}
\usepackage{todonotes}
\usepackage{xspace}
\usepackage{changepage} 
\usepackage{colortbl}
\usepackage{multirow}
\usepackage{enumerate}
\usepackage{textcomp}

\AtBeginDocument{%
  \providecommand\BibTeX{{%
    \normalfont B\kern-0.5em{\scshape i\kern-0.25em b}\kern-0.8em\TeX}}}

\setcopyright{acmcopyright}
\copyrightyear{2024}
\acmYear{2024}
\acmDOI{ENDERECO}

\acmConference[ICSE 2024]{46th International Conference on Software Engineering}{April 2024}{Lisbon, Portugal}
\acmPrice{15.00}
\acmISBN{978-1-4503-XXXX-X/18/06}

\newcommand{\rb}[1]{{\color{blue} {(\bf RB.:} #1)}}
\newcommand{\pb}[1]{{\color{red} {(\bf P.:} #1)}}

\newcommand{\code}[1]{\texttt{#1}}

\newcommand{\shortversion}[2]{#2}

\begin{document}

%%
%% The "title" command has an optional parameter,
%% allowing the author to define a "short title" to be used in page headers.
\title{Detecting Semantic Conflicts using Static Analysis}

%%
%% The "author" command and its associated commands are used to define
%% the authors and their affiliations.
%% Of note is the shared affiliation of the first two authors, and the
%% "authornote" and "authornotemark" commands
%% used to denote shared contribution to the research.
\author{Galileu Santos de Jesus}
\email{gsj@cin.ufpe.br}
\orcid{0000-0001-8749-0302}
\affiliation{%
  \institution{Centro de Informática, Universidade Federal de Pernambuco}
  \streetaddress{Av. Jorn. Aníbal Fernandes}
  \city{Recife}
  \state{Pernambuco}
  \country{Brazil}
  \postcode{50740-560}
}

\author{Paulo Borba}
\email{phmb@cin.ufpe.br}
\orcid{0000-0002-0381-2843}
\affiliation{%
  \institution{Centro de Informática, Universidade Federal de Pernambuco}
  \streetaddress{Av. Jorn. Aníbal Fernandes}
  \city{Recife}
  \state{Pernambuco}
  \country{Brazil}
  \postcode{50740-560}
}

\author{Rodrigo Bonifácio}
\email{rbonifacio@unb.br}
\orcid{0000-0002-2380-2829}
\affiliation{%
  \institution{Universidade de Brasília}
  \streetaddress{Campus Universitário Darcy Ribeiro}
  \city{Brasilia}
  \state{Federal District}
  \country{Brazil}
  \postcode{70910-900}
}

\author{Matheus Barbosa de Oliveira}
\email{mbo2@cin.ufpe.br}
\orcid{0000-0001-8758-2389}
\affiliation{%
  \institution{Centro de Informática, Universidade Federal de Pernambuco}
  \streetaddress{Av. Jorn. Aníbal Fernandes}
  \city{Recife}
  \state{Pernambuco}
  \country{Brazil}
  \postcode{50740-560}
}

%%
%% By default, the full list of authors will be used in the page
%% headers. Often, this list is too long, and will overlap
%% other information printed in the page headers. This command allows
%% the author to define a more concise list
%% of authors' names for this purpose.
% \renewcommand{\shortauthors}{Galileu Santos, et al.}
\renewcommand{\shortauthors}{} 

%%
%% The abstract is a short summary of the work to be presented in the
%% article.
\begin{abstract}
Version control system tools empower developers to independently work on their development tasks. 
These tools also facilitate the integration of changes through merging operations, and report textual  conflicts. 
However, when developers integrate their changes, they might encounter other types of conflicts that are not detected by current merge tools.
In this paper, we focus on dynamic semantic conflicts, which occur when merging reports no textual conflicts but results in undesired interference--- causing unexpected program behavior at runtime. 
To address this issue, we propose a technique that explores the use of static analysis to detect interference when merging contributions from two developers. 
We evaluate our technique using a dataset of 99 experimental units extracted from merge scenarios. 
The results provide evidence that our technique presents significant interference detection capability. It outperforms, in terms of F1 score and recall, previous methods that rely on dynamic analysis for detecting semantic conflicts, but these show better precision. 
Our technique precision is comparable to the ones observed in other studies that also leverage static analysis or use theorem proving techniques to detect semantic conflicts, albeit with significantly improved overall performance.
\end{abstract}
%%
%% Keywords. The author(s) should pick words that accurately describe
%% the work being presented. Separate the keywords with commas.
%% A "teaser" image appears between the author and affiliation
%% information and the body of the document, and typically spans the
%% page.
% \received{20 February 2007}
% \received[revised]{12 March 2009}
% \received[accepted]{5 June 2009}

%%
%% This command processes the author and affiliation and title
%% information and builds the first part of the formatted document.
\maketitle

\section{Introduction}
%\pb{A gente está usando o estilo latex recomendado na página de ICSE, com as configurações recomendadas lá? Ler em detalhe as instruções do site de submissão de ICSE. RESP.: Estamos sim, esse modelo peguei justamente do site deles. Eles falam pra usar esse template: https://www.overleaf.com/gallery/tagged/acm-official Com as seguintes configurações: sigconf, review, nonacm, anonymous acmart, discutido em: https://conf.researchr.org/track/icse-2024/icse-2024-research-track Ainda vou dar uma passada geral no formato do paper e das referências.}

%During a software development process, developers work collaboratively and independently using a central repository that controls the revisions of each program. Whenever the  teams have to contribute with development tasks, such as releasing a new feature, fixing a bug, or refactoring a specific piece of code, modifications are made to the artifacts of a system in parallel, requiring a merging process to integrate all the changes in a single version. However, issues might occur when integrating contributions from different developers, since merging might lead to a syntactically incorrect program, or even cause changes in the expected behavior of a system. Although current version control systems like Git can detect some syntactical conflicts, human intervention is still necessary for resolution, which affects the efficiency and productivity of development teams~\cite{bird2012assessing,barros2017using,da2020detecting}.

Textual (line-based) merge tools report conflicts when merging two code versions that apply changes to the same or consecutive lines.
Although common and widely studied, such \emph{textual conflicts} aren't the only problem that can occur when integrating code from two developers.

As textual merge tools focus simply on combining lines, they aren't able to detect incompatible changes that occur in areas of the code separated by at least a single line.   
So, for instance, if one developer changes a method's signature and another developer adds, lines apart, a call to the method with the original signature, textual merge tools will integrate the code with no conflicts or warnings.
The merged code, however, won't compile, characterizing a \emph{static semantic conflict}~\cite{sarma2011palantir, brun2013early, towqir2022detecting, silva2022detecting, zhang2022using,sung2020towards}.

\shortversion{,oliveira2022detecccao}{}

Even worse, textual merge tools also can't detect \emph{dynamic semantic conflict}~\cite{horwitz1989integrating, yang1992program, shao2009sca, brun2013early, pastore2017bdci, barros2017using, sousa2018verified, da2020detecting, zhang2022using}, like when the changes made by one developer affect a state element that is accessed by code changed by another developer, who assumed a state invariant that no longer holds after merging. 
In such cases, textual integration is automatically performed generating a merged program, a build is created with success for this program, but its execution reveals unexpected behavior.\footnote{We use a conflict terminology based on the language aspect that causes a conflict another common terminology is based on the process phase in which a conflict is detected.
In this one, \emph{merge} conflict corresponds to textual conflict. \emph{Build} conflict corresponds to syntactic and static semantic conflicts. \emph{Test} and \emph{production} conflicts (and undetected ones) corresponds to dynamic semantic conflicts.}
%This is caused by unintended \emph{interference} between the developers’ changes--- the behavior of the merged changes does not preserve the intended behavior of the individual changes. 
Following Horwitz et al.~\cite{horwitz1989integrating}, we put this more formally by considering dynamic semantic conflicts to be \emph{undesired interference}, where interference is defined as follows: separate changes $L$ and $R$ to a base program $B$ interfere when the integrated changes does not preserve the changed behavior of $L$ or $R$, or the unchanged behavior of $B$.
%two contributions (sets of changes) to a base program semantically conflict, that is, interfere in an unintended way, when the specifications they are individually supposed to satisfy are not jointly satisfied by the program that integrates them.

% Some researchers use different terminologies for the types of conflicts that might arise after code integration. According to \cite{kasi2013cassandra}, syntactic or textual conflicts correspond to conflicts that can be automatically identified by analyzying changes from different developers made on the same line or consecutive lines, referred to as merge conflicts. On the other hand, static semantic conflicts occur when developers are working on different branches and rename or modify the signature of a method or when there is no merge conflict, but the build process fails. These conflicts lead to the so called build conflicts~\cite{da2020detecting}\rb{pouco antes chamamos esses conflitos de ``semantic conflicts'', para em seguida classifica-los como ``build conflicts''. Acho que seria melhor definirmos esses conflitos diretamente como ``build conflicts'' RESP.: foi a definição tirada de Léuson, essa classificação foi feito por Paulo, podemos discutir isso}. Dynamic semantic conflicts, on the other hand, occur during the integration process when failures are identified through dynamic analyses (tests) or when failures are identified in production. Such conflicts are referred to as test conflicts or production conflicts.

As dynamic semantic conflicts, hereafter simply semantic conflicts, can negatively impact development productivity and the quality of software products, researchers~\cite{horwitz1989integrating, sousa2018verified, da2020detecting} have proposed techniques to detect them.
In fact, as developer's desire (specification) is hard to capture and is often not available for automated tools, these techniques simply try to detect interference.
The techniques based on theorem proving~\cite{sousa2018verified} and static analysis with system dependence graphs (SDGs)~\cite{binkley1995program, barros2017using} are computationally expensive.
%Previous research has shown that techniques based either on theorem proving~\cite{sousa2018verified} or
%static analysis with system dependence graphs (SDGs)~\cite{binkley1995program,barros2017using} are computationally expensive.
%voltei isso aqui porque os trabalhos nao trazem evidencia de que sao caras. a gente que esta argumentando isso 
Although an existing technique based on dynamic analysis~\cite{da2020detecting} has been proven less expensive, it suffers from low recall.  
%% The techniques based on theorem proving~\cite{sousa2018verified} and static analysis with system dependence graphs (SDGs)~\cite{binkley1995program, barros2017using} are computationally expensive.
%% The technique based on dynamic analysis~\cite{da2020detecting} is less expensive but suffers from low recall.  

This motivates us to explore a new alternative technique for approximating interference when merging changes made by two developers: we propose the use of lightweight static analysis on the merged version of the code, which we annotate with metadata indicating instructions modified or added by each developer.
Two of the analyses we implement (Interprocedural Data Flow and Program Dependence Graph) try to find data and control flows between instructions changed by one developer to instructions changed by the other. 
When they do find, they report interference.
The other two analysis (Interprocedural Confluence and Interprocedural Override Assignment), as explained latter, try to capture other aspects that are also suggestive of interference.

To evaluate our technique, with the aim of understanding whether it could be useful as the basis of semantic conflict detection, we conduct an experiment to understand how \emph{accurate} and \emph{computationally efficient} our analyses are for detecting interference.
%evaluates our technique using a dataset of 99 code integration scenarios. We detail our evaluation method in Section~\ref{sec:method}. 
We run our analyses on a dataset of 99 experimental units extracted from merge scenarios (quadruples formed by a merge commit, its two parents, and a base commit) mostly reused from three previous work~\cite{barros2017using, sousa2018verified, da2020detecting} on semantic conflicts.
%The scenarios we use in our experiment contain compiled versions of programs without dependencies and were collected from open-source Java projects. Some scenarios were reused from related works , while the remaining scenarios were mined using a specific tool, called Mining Framework. In each scenario within this dataset, both integrated versions modified the same declaration (Java class method or constructor).
%% Subsequently, the analyses were executed, and the tool's results were compared with a ground truth dataset. The execution time of each analysis was also recorded individually. Based on these results, metrics such as precision, recall, and accuracy were calculated. 
%
%The results of our empirical experiment (Section~\ref{sec:results}) show that our analyses achieve a precision of 0.43, meaning that it correctly identified interference in over 43\% of the attempted scenarios. The recall was 0.60, indicating that it accurately detected over 60\% of the cases. The F1 Score was 0.50, indicating a moderate overall performance of the model, considering the combination of precision and recall. The accuracy was 0.60, meaning that the tool correctly detected over 60\% of the cases. When compared to related works, we achieved better performance in terms of F1-Score and recall metrics, as well as the number of executed scenarios. \cite{silva2022detecting, oliveira2022detecccao, barros2017using, de2003breaking}.
%

Our experiment results show that our analyses have significant interference detection capability, but with potentially significant costs for handling false positives (analysis incorrectly reports interference).
We compare that with the accuracy results from previous work, running experiments with subsamples of scenarios common to our work and previous work that contains an evaluation.
Our technique shows much better F1 score and recall than the dynamic analysis technique, but with much worse precision.
Our precision is comparable to the ones presented by the SDG and theorem proving techniques; due to the experimental design of these other studies, we can't compare recall measures.
We also manually analyze and classify our analyses' false positives and negatives, and discuss how they could be mitigated. 
The performance results show that our analyses should have much better performance than previous techniques.
%
%\rb{N\~{a}o acho essencial, mas adicionaria alguns bullets refor\c cando as contribui\c c\~{o}es do trabalho.\ldots Altogether, the main contributions of this paper are:}. 

\section{Motivating Example}\label{sec:motivating}

To illustrate the concept of semantic conflict  and how our static analysis technique is capable of detecting them, consider the Merge commit code in Figure~\ref{fig:codigo-motivador}.
It integrates changes made by two developers, say Left and Right, to a Base commit. 
In Line~6, highlighted in yellow, we see Left's change. 
In Line~8, highlighted in green, we see Right's change. 
The rest of the code originates from the Base commit, which is the most recent common ancestor of the two integrated versions. 
As the changes made by Left and Right were not in the same or consecutive lines, the textual merge was successful, resulting in the illustrated code with no merge conflict or warning. 
\lstset{
  language=Java,
  basicstyle=\ttfamily,
  keywordstyle=\color{blue},
  moredelim=[il][\color{white}\bfseries]{\#},
  moredelim=[is][\color{red}\bfseries]{@}{@}
}
\begin{figure}[!ht]
    \centering
    \caption{Semantic conflict example. We highlight one developer's change in yellow and the other's in green.}
    \label{fig:codigo-motivador}
    \begin{adjustwidth}{2em}{2em}
    \begin{lstlisting}[escapechar=/, backgroundcolor=\color{white}, numbers=left]
class Text {
  String text;
  int words, spaces, comments, fixes;

  public void cleanText() {
    if (text != null  /\colorbox{yellow}{\&\& hasWeaselWords())\{}/
      /\colorbox{white}{removeWeaselWords();}/
      /\colorbox{green}{removeDuplicateWords();}/
    }
  }
  ...
    \end{lstlisting}
    \end{adjustwidth}
\end{figure}

The resulting code declares the \code{Text} class, which has a field named \code{text}, representing the string associated with objects of this class. 
It also has the \code{cleanText} method, which was originally responsible for removing weasel words (like \emph{may}, \emph{possibly}, etc.) from \code{text}, provided it was not \code{null}.
To improve text cleaning, Right added a call to the \code{removeDuplicateWords} method, which removes duplicate consecutive words from \code{text}. 
Unaware of Right's change, Left independently changed the \code{if} condition; it now checks for the presence of weasel words before trying to remove them. 

Although these changes don't result in textual conflicts, we have a problem: Right won't be happy with the resulting method behavior.
Contrary to Right's intent, duplicated words will only be removed when the text contains weasel words.
For instance, running the Base version of the method with the text \code{``the\textvisiblespace the\textvisiblespace dog''} results in the same string, but running Right's version with the same text results in \code{``the\textvisiblespace dog''}, reflecting Right's intent. 
However, running the merged method in the figure results in \code{``the\textvisiblespace the\textvisiblespace dog''}, as the input string has no weasel words.
Following the interference definition we give in the introduction, this shows that Right's behavior change is not fully preserved by the merge process, and is evidence that Left's change undesirably interferes with Right's change, causing what we call a semantic conflict.

Both developers effectively carried out their tasks in an independent way.
Each version, Left and Right, individually fulfills the requirements or specifications they were supposed to satisfy. 
However, due to unintended interference, the merged code does not satisfy the specification Right correctly implemented. 
That's why semantic conflicts of this nature are often complex and costly to identify and resolve~\cite{da2020detecting, towqir2022detecting, silva2022detecting}. 
This directly affects the quality and outcome of the products, as the final result is not as expected. 

Resolving these conflicts may require careful manual investigation, which can impact productivity. 
If not detected immediately after integration, addressing these conflicts can become even more challenging as it involves reconciling semantic and behavioral incompatibilities. 
In our example, it would be necessary to investigate whether the issue lies within the individual implementations of Left and Right or in interference between them. 
This would require a thorough investigation that goes beyond the abstraction boundaries established by the methods that \code{cleanText} calls.

\section{Detecting Interference using Static Analysis}\label{sec:solution}

To detect interference and help mitigate the negative consequences of semantic conflicts, we propose a technique that runs static analyses on the merged version of the code, which we annotate with metadata indicating instructions modified or added by each developer.
In particular, by running a static \emph{control dependence analysis}~\cite{ferrante1987program, horwitz1989integrating} on the example illustrated in the previous section, a tool could report that a method invoke instruction added by Right is control dependent on a conditional instruction modified by Left, warning about the existing  interference.  

% We explore the use of static analysis to detect interferences
% between developers' contributions in merge scenarios. Our technique
% takes as input (a) a JAR file with the compiled version of a Java program in a specific
% merge version and (b) a set of tuples where each tuple indicates the Java class name and the specific lines included or
% modified in the source code by the Left and Right parent commits.
% Identifying the lines of code in a class that either the Left or Right developer
% modified in the parent's commits is necessary to determine the entry points of our analyses
% and correctly classify the instructions as coming from a Left or Right contribution
% \rb{(in the Jimple intermediate representation) RESP.: Talvez fique muito pesado já falar em representação Jimple aqui não?}. After runnning the analyses,
% our technique outputs the possible scenarios of interference. 

We implement and rely on four analyses we describe in the following: Interprocedural Data Flow, Interprocedural Confluence, Interprocedural Override Assignment, and Program Dependence Graph, which includes the control dependency that detects the interference illustrated in the previous section.
They all try to explore likely interference situations by keeping track of the changes developers make and how they affect state elements such as global variables, object fields, system files, variables that hold method return values or raised exceptions, etc. 
We consider field accesses like \code{o.a} and \code{o.b} to refer to different state elements, even if \code{o} holds the same reference; the variable \code{o} also corresponds to a different state element.

\subsection{Static Analysis Algorithms}

For each of the four analyses we propose here, we implemented the algorithms targeting the Java programming language and took advantage of the Soot framework~\cite{vallee2010soot} as the underlining infrastructure for Java program analyses. 
In this section, we present the algorithms and how we use them to detect interference.

%\todo[inline]{RB.: Eu colocaria uma \'{u}nica implementa\c c\~{a}o da classe Text, com todos os m\'{e}todos e
%  atributos. Em cada an\'{a}lise, poder\'{i}amos referenciar os m\'{e}todos de interesse. Me incomoda um pouco
%  um m\'{e}todo com o nome countDupWords n\~{a}o retornar nada. Caso countDupWords remova palavras duplicadas
%e atualiza o atributo, um nome mais adequado seria removeDupWords.}
%\pb{countDupWords era só para contar mesmo e atualizar words, não remover palavras duplicadas.}

\subsubsection{Interprocedural Data Flow}

Consider the example in Figure~\ref{fig:df-inter}, showing the implementation of the \code{cleanText1} method of the \code{Text} class (Figure \ref{fig:codigo-motivador}). 
This method is responsible for removing extra spaces, comments, and duplicate words from the string in the \code{text} field. 
Line~12 is a change made by one developer (Left), while Line~14 is a change from another developer (Right). 
In Line~12, the call to the method \code{normalizeWhiteSpace} might lead to a \emph{definition} of the \code{text} field, as it is changed during method execution when it contains consecutive duplicate whitespace. 
On the other hand, in Line~14, the call to the method \code{removeDuplicateWords} executes an instruction that \emph{uses} the \code{text} field, as it has to be checked for duplicate words. 
These two changes lead to a \emph{def-use} data flow relationship when there are no comments in \code{text}; in this case they will involve both the definition and the use of the same state element (\code{text}). 
\begin{figure}[!ht]
  \centering
  \caption{Data Flow (DF) example based on~\cite{da2020detecting}.}
  \label{fig:df-inter}
  \begin{adjustwidth}{2em}{2em}
  \begin{lstlisting}[escapechar=/, backgroundcolor=\color{white}, numbers=left, firstnumber=11]
void cleanText1(){
  /\colorbox{yellow}{normalizeWhiteSpace();}/
  /\colorbox{white}{removeComments();}/
  /\colorbox{green}{removeDuplicateWords();}/ 
}
  \end{lstlisting}
  \end{adjustwidth}
\end{figure}

Here we characterize these \emph{def-use} relationships involving Left and Right contributions as \emph{Data Flow} (DF).
They are important for interference detection because definitions changed or inserted by one developer might affect the behavior of uses changed or inserted by the other developer.
For example, consider the execution of the base version of \code{cleanText1} with \code{text} set to \code{``the\textvisiblespace the\textvisiblespace\textvisiblespace  dog''},  with two spaces between \code{the} and \code{dog}.
It results in the same string.
Contrasting, running Left's version results in \code{``the\textvisiblespace the\textvisiblespace dog''}, as the extra space is removed.
However, running the merged version results in \code{``the\textvisiblespace\textvisiblespace  dog''}, given that the \code{removeDuplicateWords} method careless removes the second \code{the} in the input string, leaving the original space between the two duplicates, breaking Left's intention of producing a string without duplicate whitespace.
We say then that Right's change interferes wih Left's.

%We benefit from the data flow analysis literature~\cite{dfa-book,ppa-book} to identify interference. 
To implement DF interference detection, we integrate a Sparse Value Flow Analysis (SVFA) implementation that borrows from~\cite{sui2016svf} the rules for creating a sparse graph that allows us to find interprocedural paths from \emph{defs} to \emph{uses} involving a given state element. 
SVFA is often used in the context of taint analysis~\cite{sui-tse-2014,shi-pldi2018} to identify source-sink paths~\cite{arzt-pldi-2014,grech-oopsla-2017}. 
So we run it twice: the first execution considers Left and Right contributions as source and sink statements, respectively; the second execution is the inverse. 
This way we cover data flows from Left to Right contributions and vice versa.
In the first execution, we mark ``definitions'' in the Left contributions as \emph{source} vertices and ``uses'' from the Right contributions as \emph{sink} vertices in the graph. 
In the most basic case, we can define a data flow edge in the SVFA graph as one that connects a vertex where a variable is defined to another vertex where the same variable is used. 
More complex cases deal with the flow of data when binding actual and formal arguments in method calls, return statements, and field stores and field loads.
We report interference when there is a path in the graph that starts with a source vertice and ends with a sink vertice.

%As a matter of fact, we execute the algorithm for detecting Interprocedural Data Flow twice for every merge scenario. The first execution considers Left and Right contributions as source and sink statements, respectively. The second execution is the inverse; Left contributions are sink statements, while Right contributions are source statements. This way, we can cover data flows from Left to Right contributions and vice versa.

\shortversion{\pb{Detalhes importantes que podemos falar se sobrar espaço: como é feita a marcação? durante a análise, visita-se chamadas de métodos não marcadas? por que? SVFA é X-sensitive?}}{}

%% \todo[inline]{RB.: Acho que Paulo sugeriu remover esses detalhes do grafo. RESP.: Isso mesmo, sugerindo que todos já sabem, porém deixar como implementamos e quais características.}
%% The SVFA graph that \ssm builds for the example in Figure \ref{fig:df-inter} is illustrated in Figure \ref{fig:df-grafo-inter}. Here, we can see that there is a path from the yellow vertex (from the method \code{normalizeWhiteSpace()}) to the green vertex (from the method \code{removeDuplicateWords()}), indicating an interference between the two contributions. Note that this graph was presented in a simplified form, without including nodes for interprocedural method calls, in order to simplify the example. However, our implementation follows this technique, with a larger graph that encompasses more methods.

%% \begin{figure}[htbp]
%%     \centering
%%     \caption{Example of a Data Flow (DF) graph for the Figure \ref{fig:df-inter}.}
%%     \label{fig:df-grafo-inter}
%%     \includegraphics[width=8cm]{images/df_example.jpg}
%% \end{figure}

\subsubsection{Interprocedural Confluence}

Interference can also occur when there is no data flow between the integrated changes, but there are data flows from the integrated changes to a common statement; this statement behavior might be affected by both changes, characterizing interference.
The \code{Text} method in Figure~\ref{fig:cf-inter} illustrates that.
The \code{countFixes} method returns the sum of the number of needed changes stored in the \code{spaces} and \code{words} fields (see Figure \ref{fig:codigo-motivador}).
Left inserted Line~17, and Right inserted Line~19. 
These inserted method calls simply modify different state elements (\code{spaces} and \code{words}), so there is no data DF between them.
However their state changes flow to the \code{return} statement on Line 20. 
In this case, the result of a call to the method \code{countFixes} will be different when considering the Left version, the Right version, and the Merge version, characterizing interference. 
\begin{figure}[!ht]
    \caption{Confluence Flow (CF) example.}
    \label{fig:cf-inter}
    \begin{adjustwidth}{2em}{2em}
    \begin{lstlisting}[escapechar=!, backgroundcolor=\color{white}, numbers=left, firstnumber=16]
int countFixes(){
  !\colorbox{yellow}{countDupWhiteSpace();}!
  !\colorbox{white}{countComments();}!
  !\colorbox{green}{countDupWords();}!
  !\colorbox{white}{return}!!\colorbox{red}{spaces + words}!;
}
    \end{lstlisting}
    \end{adjustwidth}
\end{figure}

To detect \emph{Confluence} (CF), we execute our interprocedural data flow analysis
algorithm (SVFA) twice, once marking Left statements as \emph{source} and
base statements as \emph{sink}, and once marking Right statements as \emph{source} and the
base statements as \emph{sink}. 
The goal is to determine if there are \emph{definitions} from Left and Right that assign expressions to state elements flowing to Base statements that \emph{use} these state elements.

\subsubsection{Interprocedural Override Assignment}

Differently from the previous two analyses, which focus on detecting data flows between the integrated changes or from them to a common statement, here we capture situations where state updates from one developer are overridden by the other.
Interference occurs in such cases because the behavior of the changes from one developer is not preserved in the merge version.
%are     Occurs whenever the contributions from two developers (working in the Left and Right parents of a merge scenario)
%assign different values to the same state element. The interference happens because the changes coming from different branches
%to the same state element can alter the final result of a computation.
%In other words, one developer (Left) may expect a state element (such as a variable)
%to be assigned to a certain value, but, in the end, such a value might have been overridden by the contributions
%from the Right developer.
%
Figure~\ref{fig:oa-inter} illustrates that. 
Left inserted the call to the \code{countDupWhitespaces} method, while Right inserted the call to \code{countDupWords}. 
Both methods simply set the instance field \code{fixes} with the number of needed fixes they counted, whereas \code{removeComments} sets another field. 
%In this case, the contributions from Left and Right interfere with each other, as they modify the same state element in different parts of the code. Such interference can influence the final outcome of the program, since the since the Left and Right contributions might not preserve the expected outcomes from both developers.
%
% se incrementasse, ao inves de setar, seria interferencia mas nao conflito!
%
Because of Right's change, the count of the number of duplicate whitespace is then lost in the merge revision, characterizing the interference.

To implement \emph{Override Assignment} (OA), we use an interprocedural algorithm that records definitions in two sets: $S_l$ contains the definitions coming from the analysis of code annotated as Left, whereas $S_r$ contains the definitions coming from the Right.
As soon as we analyze Base code that defines a state element in one of the sets, we remove this element from the set.
We report interference when we try to record in one of the sets a definition of a state element that is already in the other set, or when there is a containment relationship between the elements: \code{o} is in one of the sets and we try to record \code{o.a}, or vice-versa.
\begin{figure}[!ht]
  \centering
  \caption{Override Assignment (OA) example.}
  \label{fig:oa-inter}
  \begin{adjustwidth}{2em}{2em}
  \begin{lstlisting}[escapechar=!, backgroundcolor=\color{white}, numbers=left, firstnumber=22]
void generateReport() {
  !\colorbox{yellow}{countDupWhitespaces();}!
  !\colorbox{white}{countComments();}!
  !\colorbox{green}{countDupWords();}!
}
}//end of the Text class
  \end{lstlisting}
  \end{adjustwidth}
\end{figure}

\subsubsection{Program Dependence Graph}

The last analysis we discuss is illustrated in our motivating example (see Sections~\ref{sec:motivating}).  
% in  described Recall the code snippet of Figure~\ref{fig:codigo-motivador}. In that case, 
% the Left contribution modifies the \code{cleanText()} method on Line 6, adding the conditional
% statement \code{if (text != null \&\& hasWeaselWords())}. Whilst, the Right contribution modifies the same method (\code{cleanText()}) on Line 8,
% adding a call to the \code{removeDuplicateWords()} method. In this case,
% a \emph{control dependence} occurs because the Left contribution (Line 6)
% might interfere with the control flow of the program that is necessary to reach
% the contribution from Right (Line 8).
%
% We say that the body has a control dependence, meaning that in order for the
% internal commands of the conditional to be executed, the execution flow must
% pass through it. Consequently, the right developer made a modification within
% the body of the conditional, which, in order to reach it, must necessarily pass
% through the change made by the other developer, resulting in a direct interference in its execution.
There, however, we mention only the \emph{control dependency} aspect. 
But our implementation builds a Program Dependence Graph (PDG)~\cite{ferrante1987program, horwitz1989integrating}; the program instructions are vertices, while the edges represent either
control or data dependencies.
Due to the control dependencies, we detect interference in the motivating example. 
%, observing that (a) a data dependence arises when the result of one instruction is used as input to another instruction and (b)
%a control dependencies occurs when the execution of one instruction depends on some condition of another statement~\cite{horwitz1989integrating}.
We build our graph using similar rules to Ferrante et al.~\cite{ferrante1987program}. 
We generate a PDG for each method under analysis, and report interference when there is a path (having control or data dependencies) from a Left instruction to a Right instruction, or vice-versa.
The existence of a path indicates suggests that the changes made by one developer might have an influence on the execution of changes made by the other. 

% So, considering the code snippet of Figure~\ref{fig:codigo-motivador}, we build a
% PDG having two auxiliary vertices representing the entry and return points of a method. The entry point represents the initial node, and the return point represents the final node.
% We annotate every other remaining vertices (the program statements vertices) to identify if
% it belongs to the changes coming from Left or Right. Figure \ref{fig:cd-graph} illustrates the PDG
% for the example in Figure~\ref{fig:codigo-motivador}. In this case, the yellow vertice corresponds to a Left contribution,
% and the green vertice corresponds to a Right contribution.  It is possible to see a control dependence from Left to Right, representing that
% the Left contribution (Line 6) might influence the execution of the Right contribution (Line 8).

% \rb{Esse \'{e} um grafo PDG??? RESP.: É sim, é um grafo seguindo as definições de horwitz1989integrating, nesse caso não temos nenhum fluxo de dados, apenas Control Dependence}

% \begin{figure}[htbp]
%     \centering
%     \caption{Representation of a PDG for Figure \ref{fig:codigo-motivador}.}
%     \label{fig:cd-graph}
%     \includegraphics[width=9cm]{images/cd.pdf}
% \end{figure}

\subsection{Limitations}

Our analyses suffer from limitations that are typical in static analyses. 
First, we do not detect interference that might result from the execution of code that uses Java reflection, and we do not analyze Java native methods. 
Second, our analyses do not differentiate specific positions of an array. 
%That is, when the state element is just a position of an array, we consider the entire array in the subsequent analysis. 
We also have to compromise precision with the analysis complexity and execution time. 
For instance, our current implementation for detecting OA is not \emph{object-sensitive}. 
This could lead the analyses to report interference that does not exist (a false positive). 
Similarly, the current SVFA implementation we use is not fully accurate (as expected) and might lead to false positives and false negatives. 
This is particularly true for corner cases whose state elements involve fields.

%% To better understand, let's consider the example code in Figure \ref{fig:codigo-motivador}, where we generate the Control Dependence graph in Figure \ref{fig:cd-graph}, which is equivalent to the PDG since there are no additional edges in this case. Intuitively, we generate a PDG graph that represents the flow of information through a program and check if there is a path between contributions using their control and data dependence edges.

%% When there is a data dependence, the output of one instruction is used as input for another. Control dependencies occur when the execution of an instruction depends on a condition defined in another instruction. Order dependencies occur when the definition of a variable is used in another instruction before its own definition. By identifying these dependencies, we can understand how the program instructions interact with each other and consequently identify possible interferences that may occur between them. For example, if two instructions have a control dependence, it is possible that one of them is preventing the execution of the other, causing a bottleneck in the program.

\section{Evaluation Method}\label{sec:method} 

\begin{figure*}[h]
    \centering
    \caption{Experiment setup. We start with open-source Java projects from GitHub, and select merge scenarios with methods or constructors changed by both developers. For each scenario, we build the merge commit version, generating a JAR file without external project dependencies. We also extract information about the method changed by both developers, and the specific lines they have changed; this method is given as the analyses entry point, and the line information is used to annotate the internal code representation adopted by our analyses. Finally, we run the implemented static analyses, collect execution time information, and compare interference ground truth with the logical disjunction of the four analyses results.}
    \includegraphics[width=\textwidth]{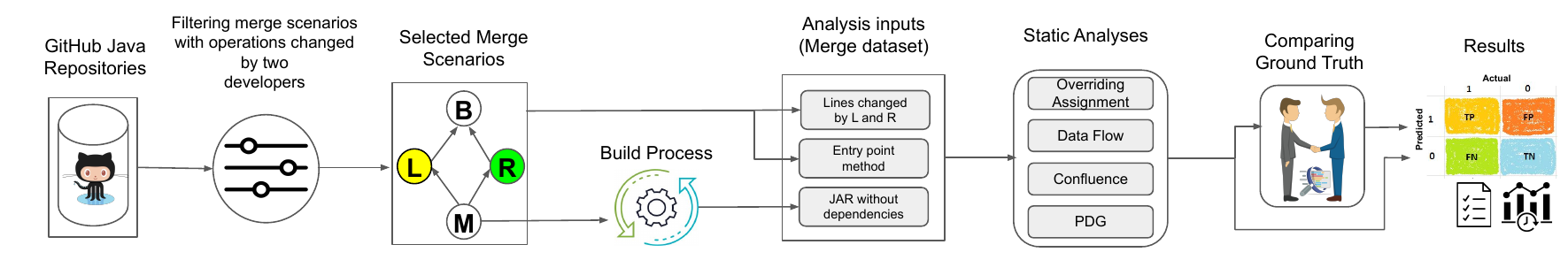}
    \label{fig:methodology}
\end{figure*}

To evaluate the potential of our static analyses for detecting interference, we follow the experimental design illustrated in Figure~\ref{fig:methodology} and explained in the rest of this section.
We want to understand how \emph{accurate} and \emph{computationally efficient} our analyses are for detecting interference, with the aim of understanding whether they could be useful as the basis of semantic merge conflict detection tools.
\subsection{Merge Scenario Selection}

We start with open-source Java projects from GitHub, and select merge scenarios (quadruples formed by a merge commit, its two parents, and a Base commit) having operations (methods or constructors) that were independently changed by both developers.
We rely on a number of Java projects and scenarios that appeared in previous studies on semantic merge conflicts~\cite{barros2017using, sousa2018verified, da2020detecting, silva2022detecting}, and mine new ones following similar criteria. 

% The new scenarios we mine come from a list of 2 (two) projects that we selected... \pb{DONE Como escolhemos a lista de projetos dos cenários que mineramos? E por que? RESP.: São 3 novos CASOS, O projeto storm está em Léuson, porém foi minerados dois novos CASOS: cenários 9 e 10 de storm com hash (ad2be678831b3b060229fd936e3908110162b7ac). O projeto MPAndroidChart está em Sousa, porém foi minerado um novo CASO com o mesmo hash (af114d180da6ec5633d32c701ff5467f7629fcf3) para o método onProgressChanged}.  
% We mine these projects from X/Y/ZZZZ to X/Y/ZZZZ... \pb{Que período mineramos? E por que? RESP.: não diz nada a respeito da data de mineração
% O projeto storm foi gerado o . jar em Apr 22, 2020. 
% O projeto MPAndroidChart foi gerado o .jar em  Jun 13, 2020.
% Ambos minerando e executando o mining neste período.
% Datas dos commits para o repositório apache/storm:
% Data: 2017-02-01T21:28:00Z
% Datas dos commits para o repositório PhilJay/MPAndroidChart:
% Data: 2016-06-26T20:04:19Z
% Data: 2016-06-26T20:04:19Z
% Data: 2016-06-26T20:04:19Z
% Data do commit mais antigo do dataset: 2013-10-27T05:53:47Z
% Data do commit mais recente dos novos cenários: 2017-02-01T21:28:00Z}
%\pb{Descartamos algum ou pegamos todos os merge commits que atendiam aos critérios nesse período?} 

We discard fast-forward and criss-cross merge commits~\cite{chacon2014pro}.
The former don't correspond to code integration from \emph{two} developers, the phenomenon we want to investigate; 
the latter don't have a single most recent common ancestor, therefore not characterizing a single merge scenario.
Following previous work, we focus on scenarios with declarations changed by both developers, as, in this way, one can more easily establish interference ground truth. 
Contrasting with previous work on dynamic analysis~\cite{da2020detecting, silva2022detecting}, we discard scenarios having only \emph{field} declarations changed by both developers;
this is also adopted by related work on information flow analysis~\cite{barros2017using} and verification~\cite{sousa2018verified} since they also require a \emph{method} or \emph{constructor} declaration as the analysis entry-point.
%\pb{Galileu, o critério de Sousa inclue atributo tb? RESP.: Não. Ele pega somente alterações no mesmo método. "In particular, to reduce the scope of our problem, we focus in interference caused by developers same-method contributions."} 

%The data mining tool, \emph{miningframework}, takes a CSV file as input, which contains a list of repositories on GitHub. Using this information, forks of the listed repositories are created to set up a Continuous Integration (CI) tool. Subsequently, the framework generates the merge scenarios, which include the files from the base, left, and right versions

As a merge scenario might have more than one operation that was independently changed by both developers, we consider pairs $(scenario,operation)$ as our \emph{experimental units}. 
We mostly reuse units that originally appeared in the datasets of previous work: 35 units from~\cite{silva2022detecting}, 31 from~\cite{barros2017using}, and 30 from~\cite{sousa2018verified}, as illustrated in Table~\ref{tab:resumo-mergedataset}.
Together with the three new units that first appear in this work, we have a total of 99 experimental units, associated with 54 merge scenarios extracted from 39 projects.
\begin{table}[h]
    \centering 
    \caption{Summary of reused experimental units. Second column shows the number of units we reused in our dataset, whereas the third column shows the number of units that originally appear in the dataset of the corresponding source.}
    \label{tab:resumo-mergedataset}
    \begin{tabular}{c c c}
        \cellcolor{gray!10}\textbf{Source} & \cellcolor{gray!10}\textbf{Our Dataset}  & \cellcolor{gray!10}\textbf{Original in  
        paper} \\ \hline
        \cite{silva2022detecting} & 35 & 42 \\ \hline 
        \cite{barros2017using} & 31 & 35\\ \hline 
        \cite{sousa2018verified} & 30 & 52\\ \hline 
    %   Mined & 3 \\ \cline{1-2}
    %   \cellcolor{gray!25}\textbf{Total} & \cellcolor{gray!25}\textbf{99} \\ \cline{1-2}
    \end{tabular}
\end{table}

As our experiment requires entry points and the creation of JAR files for each unit, we discard from the related work datasets the experimental units with build issues\footnote{Units with compilation problems, need for installing extra frameworks, generated JAR without method to be analyzed, or several JARs but none conforms to the name conventions adopted by our experimental infrastructure.} or no operation changed by both developers; so the differences in numbers appearing in the second and third columns of Table~\ref{tab:resumo-mergedataset}.
All the information (commit hashes, source project and class, summary of changes made by each developer, etc.) about the projects, scenarios, and units is available from our online appendix~\cite{online-appendix}. 
Among the projects, we have \href{https://github.com/elastic/elasticsearch}{elasticsearch} and \href{https://github.com/webbit/webbit}{webbit}, showing a considerable degree of diversity concerning dimensions such as project domain (the first is a search engine, the second is an HTTP server), size (1096 KLOC versus 20 KLOC), stars (4,232 versus 68,509), commits (69,697 versus 770), forks (23,351 versus 186), and number of collaborators (2,054 versus 39).

\subsection{Preparing Analyses Inputs}

For each scenario selected in the previous section, we build the merge commit version.\footnote{As we don't analyze the other commits in a scenario, there is no need to build them.}
The resulting JAR file is used as input for our static analysis infrastructure, which translates Java bytecode into the Jimple intermediate representation (one of the
representations the Soot framework supports)~\cite{vallee2010soot}, before running the analyses.
% \pb{Matheus, nenhum dos JARs gerados usou o compilador do Eclipse?}
% Resp: Nenhum dos .jar foi gerado com eclipese. Fizemos apenas um teste em um único cenário, para verificar a diferença mas não foi incorporado no mergedataset
We generate JAR files without external project dependencies. 
This improves analysis performance, as less code is analyzed. On the downside, this likely degrades analysis accuracy.
As this trade-off seems promising for potential practical uses of static analysis for detecting interference, we opted for this experimental design.

Besides a JAR file, our analysis infrastructure requires two more inputs.  
So, for each merge scenario we also extract information about the operation (class method or constructor)
changed by both developers, and the specific lines they have changed, that is, included or modified; pure deletions are not considered.
The operation is given as the analyses entry points.
In case more than one operation was changed by both developers, the scenario yields more than one experimental unit.

The line information is used to annotate the internal code representation adopted by our analyses, which, for example, detect interference by finding data and control flows between instructions changed by one developer to instructions changed by the other. 
So knowing which instructions were changed by each developer is essential for our analyses to work.
We obtain the lines changed by one parent by taking its syntactical diff with the Base commit. For that, we use the DiffJ tool.\footnote{DiffJ compares Java files based on their code, available at: \href{https://github.com/jpace/diffj}{https://github.com/jpace/diffj}} Comparing to a textual diff, DiffJ brings direct and more accurate information about the changed declarations and the lines changed inside them. 
Using the extracted line information, our infrastructure annotates the Jimple representation of the code.

\begin{figure*}[htbp]
    \centering
    \caption{Realistic experimental unit example. Original code on the left, refactored one on the right. Commit \href{https://github.com/elastic/elasticsearch/pull/13520/commits/f3d63095dbcc985e24162fbac4ee0d6914dc757d}{0d6914dc757d} from project \emph{elasticsearch}.}
    \label{fig:realistic-example}
    \includegraphics[width=\textwidth]{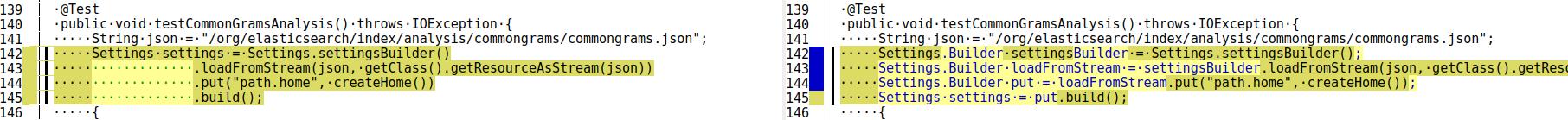}
\end{figure*}

As two of the 99 selected experimental units have entry points containing long chains of method calls that spread across a number of lines, they were slightly refactored before building their JAR files.
We manually extracted local variables that are initialized with subexpressions of the chain, in such a way that the original behavior is preserved, but changed line information is more accurately carried on to the bytecode and the intermediate representation used by our analyses. 
We illustrate this in Figure~\ref{fig:realistic-example}, showing the original code on the left and the refactored one on the right.
In the original code, line~144 corresponds to a change made by one developer, say Left; line~143 is a change made by developer Right.
As the same statement (the assignment starting on line~142) contains changes made by both developers, incorrect line information might reach the analyses.
In fact, the method call on line~144 is translated to instructions that first store on an array the \texttt{put} arguments, and then invoke this method.
The array storage instructions, though, are incorrectly annotated as Right, leading to incorrect analysis results indicating that there is a dataflow from Right to Left. 
With the refactored example on the right, each statement is on a different line, leading to accurate annotation and analyses results. 
Similarly, we manually refactored a ternary if expression that was spread across two lines. 
So we can say our sample consists of 96 real scenarios and three realistic scenarios.

%\pb{Foi exatamente isso? Em todos os 3 casos? RESP.: Foram 3 cenários: 24 (refatoração em um if ternário, subiu uma linha, pois era feito com uma quebra de linha), 90 e 91 (ambos com extract variable em um build que é feito em uma linha com quebra)}

For 31 merge scenarios, we could not automatically build the project with the existing build configuration files.
In these cases, we manually changed the configuration files, for example by updating dependencies and removing test execution steps in the build process. 
In most cases, we had to simply change the configuration file to properly use the Maven Assembly plugin~\cite{maven-plugin}.
In two cases we had to manually change the code to successfully build the project.
In one of them we just commented part of a test code that was breaking the build due to the lack of external resources.
In the other we had to refactor the code to work with a newer version of a dependency, as it had changed a constructor interface.
For helping replication, all modifications are documented in README files available in our  repository with the whole dataset (see our online appendix~\cite{online-appendix}). 

%The entire process is available in the \emph{mergedataset} repository, which containss the collected merge scenarios generated by the tool\footnote{Repository with the collected merge scenarios generated by the tool, available at: https://github.com/spgroup/mergedataset}.

\subsection{Running Analyses}

With the prepared analyses inputs, in this step we run, for each experimental unit, the algorithms we implemented to detect interference (Section~\ref{sec:solution}). 
For performance reasons, the interprocedural analyses (that detect Data Flow, Confluence, and 
Overriding Assignments interferences) are all executed with nesting level five, meaning that they only analyze methods called up to five levels deep from the entry point. 

These analyses are also configured to use the Spark~\cite{spark-paper,lhotak2003spark} points-to analysis framework, increasing the call graph precision---in comparison to
other alternatives like Rapid Type Analysis and Class Hierarchy Analysis~\cite{PGL-014}. 
We also configured Soot so that it keeps line numbers obtained from bytecode and does not consider classes in standard Java libraries, except for basic classes like \texttt{String}, \texttt{RuntimeException}, and the ones that simply encapsulate primitive types (\texttt{Integer}, \texttt{Long}, etc.).
So classes like \texttt{ArrayList} and \texttt{HashMap} are not considered, for example.

Besides collecting the analysis results reporting interference or its absence, we collect execution time information for each analysis and the logical disjunction of all the analysis results.
For that, we run each analysis ten times for each of the 99 experimental units.

\subsection{Summarizing Results}

In this final step of our experiment, we compare the interference ground truth with the logical disjunction of the analyses results, and compute precision, recall, F1, and accuracy metrics.
At least two researchers manually analyze each unit and check for interference using the definition in the introduction, and conditions that imply interference~\cite{horwitz1989integrating}.\footnote{There is a state element $x$ such that Base, Left, and Right compute different values for $x$; Left (or Right) computes a different value for $x$ compared to both the Base program and the merged version; or  Base, Left, and Right compute the same value for $x$, but the merged version computes a different value.}

In case of disagreement, the researchers discuss the experimental unity with one more researcher and reach a verdict. 
In case of agreement, they present the experimental unity and its evaluation to the other researchers in the group. 
To significantly reduce the chances of misjudgment, each interference verdict comes with a manually designed test case that reveals the interference (for instance, by breaking on the Base version, passing on one of the parents version, and breaking again in the merge version). 
Similarly, each non-interference verdict has an explanation of why we could not design such a test case.

We also summarize and analyze the execution time information in a number of ways.
These activities and the ones in the previous step are automated and implemented as a docker container for reproducibility purposes
(see our online appendix~\cite{online-appendix}).\footnote{The activities in the first two steps are also automated, but require manual effort in a number of scenarios,
as explained earlier. That is why the replication container contains only the activities in the last two steps of our experiment.} 

Additionally, we manually investigate all the scenarios classified as false positives (the analyses incorrectly report interference) and false negatives (the analyses don't report existing interference).  
We classify the causes behind the analyses inaccuracies, to better explain our results and understand how it can be improved by future work.

\section{Evaluation Results}\label{sec:results}

We now report the \emph{accuracy} and \emph{computational efficiency} results we obtained by running the experiment as described in the previous section, using our sample of 99 experimental units.

\subsection{Accuracy Results}

We depict the confusion matrix of our experiment as Table~\ref{tab:resultados_experimento}, which shows that our sample contains 33 units with interference, and 66 without interference.
The logical disjunction of our analyses results--- hereafter ``our analyses,'' for simplicity--- reports 26 false positives (incorrectly predicted interference) and 13 false negatives (non reported interference).

\begin{table}[h]
    \centering
    \caption{Confusion matrix. Prediction is established by the logical disjunction of the four analyses results.}
    \label{tab:resultados_experimento}
    \begin{tabular}{l|l|c|c|c}    
        \multicolumn{2}{c}{} & \multicolumn{2}{c}{\textbf{Actual}} & \\
        \cline{3-4}
        \multicolumn{2}{c|}{} & \cellcolor{gray!10}\textbf{Positive} & \cellcolor{gray!10}\textbf{Negative} & \multicolumn{1}{c}{\textbf{Total}} \\
        \cline{2-4}
        \multirow{2}{*}{\textbf{Predicted}} & \cellcolor{gray!10}\textbf{Positive} & $20$ & $26$ & $46$ \\
        \cline{2-4}
        & \cellcolor{gray!10}\textbf{Negative} & $13$ & $40$ & $53$ \\
        \cline{2-4}
        \multicolumn{1}{c}{} & \multicolumn{1}{c}{\textbf{Total}} & \multicolumn{1}{c}{$33$} & \multicolumn{1}{c}{$66$} & \multicolumn{1}{c}{$99$} \\
    \end{tabular}
\end{table}
These numbers result in the accuracy measures we show in the first SA (for Static Analysis, that is, the technique we evaluate here) column grouped under Techniques in Table~\ref{tab:trabalho-relacionados}.
The last row in that column indicates the number of experimental units considered for computing the measures.
With 26 false positives, our analyses precision reaches only 0.43, but our recall reaches 0.60.
This shows significant interference detection capability, but with potentially significant costs for handling false positives.

To better contextualize these accuracy results, we investigate how our analyses compare with interference detection techniques  proposed by previous work, which report at least enough information for computing precision.
As comparing accuracy measures based on different datasets could introduce bias that would be hard to identify and explain, and running other people tools on a larger sample could be difficult, we recompute the SA metrics for subsamples that are common to our dataset and the datasets used by each related work.

\begin{table}[h]
    \centering
    \caption{Accuracy metrics and comparison with previous work and their datasets. We use SA for the Static Analyses we propose here, IF for the Information Flow analysis used in~\cite{barros2017using}, DA for the Dynamic Analysis used in~\cite{da2020detecting, silva2022detecting}, and VA for the Verification Algorithm used in~\cite{sousa2018verified}.}
    \label{tab:trabalho-relacionados}
    \begin{tabular}{lc|cc|cc|cc}
        \hline
        \textbf{Metrics}   &  \multicolumn{7}{c}{\textbf{Techniques}}  \\ \hline\hline
        
         & \textbf{SA} & \textbf{SA} & \textbf{DA} & \textbf{SA} & \textbf{IF}  & \textbf{SA} & \textbf{VA} \\   \hline
          
        \textbf{Precision}  & {0.43} & {0.45} & {0.80} & {0.65} & {0.54}  & {0.16} & {0.22}  \\ \hline
        \textbf{Recall}  & {0.60} & {0.60} & {0.14} & {0.88} & {-} & {0.20} & {-} \\ \hline
        \textbf{F1 Score}  & {0.50} & {0.52} & {0.24} & {0.75} & {-} & {0.18} & {-} \\ \hline
        \textbf{Accuracy}  & {0.60} & {0.58} & {0.66} & {0.67} & {-} & {0.70} & {-} \\ \hline
        \textbf{Units}  & {99} & {75} & {75} & {31} & {31} & {30} & {30} \\ \hline
    \end{tabular}
\end{table}
So the SA--DA column group under Techniques in Table~\ref{tab:trabalho-relacionados} shows the accuracy measures we obtained by running our analyses and the Dynamic Analysis (DA) proposed in~\cite{silva2022detecting}, using 75 experimental units that are common to both our dataset and the one used by this related work.\footnote{This work dataset contains 85 units, 42 of which originally appear in that paper, and 43 are reused from previous work. Of these 85 units, we could use in our experiment only 75, as four of them are related to changes in field declarations, and six we could not successfully build.}
With this subsample, our analyses precision
is at 0.45, slightly higher than with our complete dataset; we see no change on recall.
The DA precision is 0.8 (against 0.69 in their complete dataset, but not shown in the table), but recall is at 0.14 (0.32 in their dataset).
% \pb{DONE Quais os números reportados por Léuson, com a amostra dele de 85 units? RESP.: 
% para os trabalhos relacionados, temos os seguintes valores:
% Léuson: precision: 0.6923076923; recall: 0.32; F1 Score: 0.4390243902; Accuracy: 0.7294117647
% Léuson utilizando nosso LOI: precision: 0.8181818182; recall: 0.26; F1 Score: 0.3913043478; Acurracy: 0.6705882353
% Roberto: precision: 0.5714285714
% DeSouza: precision: 0.8461538462
% Utilizando os nosso cenários e LOI, temos o apresentado na tabela.
% } 
% \pb{Ajeitar os números de DA vide os ajustes que léuson fez na tabela. RESP.: Com o ajuste de Léuson, temos 2 FP e ele colocou um * em outro que não é FP, porém, um desses FP não foi inserido na comparação com os 75 units, foi o cenário AtmosphereConfig ID 117, ele está com not-found, é um cenário originalmente de Roberto.}

In the SA--IF column group we compare our analyses with the Information Flow analysis (IF) proposed in~\cite{barros2017using}.
We had only 31 common experimental units in the two datasets. 
With this subsample, our analyses precision (0.65) and recall (0.88) are much higher than with our complete dataset.
Our precision is also higher than the one presented by IF (0.54, against 0.57 in their full dataset).
Unfortunately, due to the experiment design adopted by the authors--- they only collect and manually analyze merge scenarios for which their tool reports interference--- we cannot compute recall and the other metrics for IF. 
\shortversion{\rb{Alguma ideia do porque do resultado ser melhor para esse dataset? Essa se\c c\~{a}o est\'{a} excelente.}}{}
%\pb{DONE Quais os números reportados por Roberto, com a amostra dele de 35 units? RESP.: Roberto: precision: 0.5714285714}

Finally, in the SA-VA column group we compare our analyses with the Verification Algorithm (VA) proposed in~\cite{sousa2018verified}.
We had only 30 common experimental units in the two datasets.
With this subsample, our analyses precision (0.16) is much lower than with our complete dataset, but comparable to theirs (0.22, against 0.85 with their complete dataset).
This variation is due to a subsample with only five positives, our tool missing four of them; the percentage of true positives in the sample is approximately seven times lower than in the complete sample, significantly affecting precision.\footnote{With five false positives (among 25 negatives), the percentage of false positives in the sample is comparable to the complete dataset.}
As the authors manually analyze only merge units for which their tool reports interference, their paper has no information on false negatives, and therefore doesn't report recall and the other metrics. 

\subsection*{Manual analysis of false positives and negatives}

\begin{figure*}[h]
    \centering
    \caption{Classification of our analyses false positives (incorrectly reported interference) and false negatives (non reported interference).}
    \includegraphics[width=\textwidth]{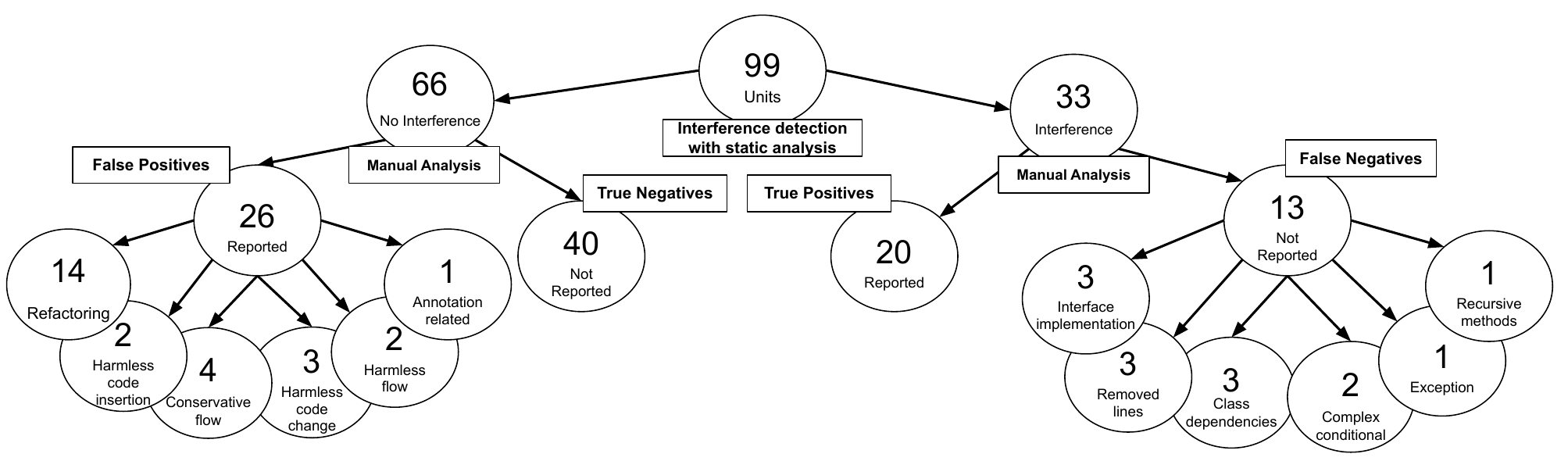}
    \label{fig:resultados}
\end{figure*}

To better understand the reasons for our analyses inaccuracies, we manually analyze the 26 false positives and 13 false negatives we show in Table~\ref{tab:resultados_experimento}.
Figure~\ref{fig:resultados} summarizes the analysis results, giving the reasons for misclassification, and the number of units misclassified in each category. 

\begin{figure*}[h]
    \centering
    \caption{Refactoring false positive. The Base version appears on the center. The parents changes appear on the left and the right. Experimental unit from project \emph{dropwizard}, merge commit \href{https://github.com/dropwizard/dropwizard/commit/ddd15a681bf42360337844412cae4aba1556eb88}{4aba1556eb88}.}
    \includegraphics[width=\textwidth]{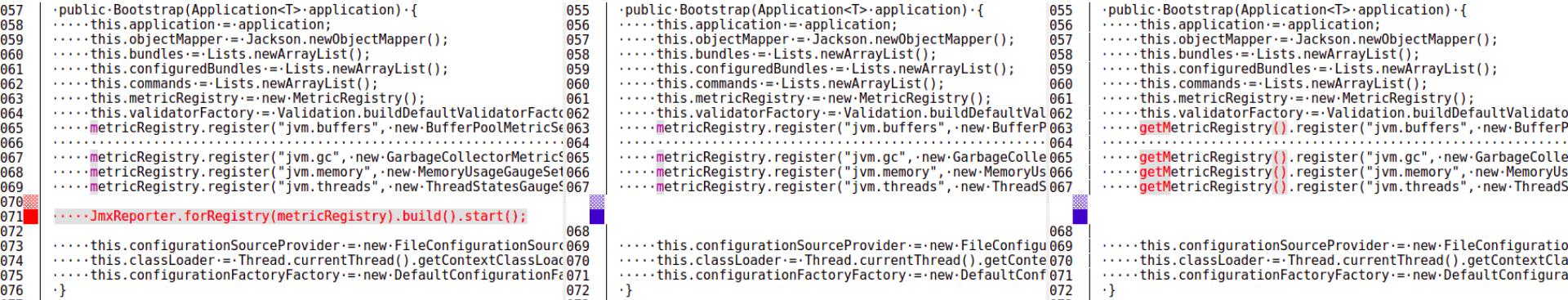}
    \label{fig:fp_example}
\end{figure*}

The main reason for false positives, with 14 misclassifications, is refactoring changes.
Figure~\ref{fig:fp_example} illustrates an example. 
The changes on the Right correspond to an extract method refactoring, where direct accesses to the \texttt{metricsRegistry} field were replaced by calls to the corresponding \texttt{get} method.
This does not affect the behavior of the illustrated method, and therefore does not interfere with the change illustrated on the left.
However, our DF analysis incorrectly reports interference because all calls to the \texttt{get} method are annotated as \emph{right}, the new call to \texttt{forRegistry} is annotated as \emph{left}, and there is a dataflow from \emph{right} to \emph{left} instructions; the \texttt{forRegistry} call reads the \texttt{metricsRegistry} field that is written by the \texttt{register} method calls.
%\pb{Que análise DF reportou nesse caso? Qual foi o fluxo? De onde para onde? RESP.: DF detectou fluxo, das linhas do getMetricRegistry() [65, 67, 68 e 69] alteradas por right por refactoring, extraindo o nome do atributo e colocando um get para ele, que escreve em metricRegistry, usado na linha 071 por left}
% Esse exemplo de FP é do caso 32
In summary, \emph{structural} refactorings\footnote{Not all refactorings are interference free. Consider code with a redundancy, like two null pointer checks. If developer Left removes one of the checks, and Right removes the other, both applied non structural refactorings, but the changes interfere.} lead to unnecessary code annotation in our technique, and that might incorrectly trigger our analyses interference detection.  

The are a number of other reasons for false positives.
We observe Harmless code insertion when both developers add new statements--- like assignments to different fields--- that change behavior but affect unrelated state elements; a new flow, though, is created due to the insertions. 
%\rb{DONE Qual an\'{a}lise espec\'{i}fica est\'{a} gerando FPs para Harmless code insertion? RESP.: PDG  em um caso devido o uso de throws, gera um exception; DF nos outros quatro casos, motivos: imprecisão da análise pois acaba indo muito a fundo e não identificando qual objeto fez a referência corretamente, mesmo caso do exemplo 17 do benchmark. Outro motivo é a marcação de linhas erradas, acaba marcando statements de forma incorreta e encontrando fluxo.}
% Esse exemplo de FP é do caso 40
% \begin{figure*}[h]
%     \centering
%     \caption{Harmless code insertion false positive. The Base version appears on the center. The parents changes appear on the left and the right. Experimental unit from project \emph{voldemort}, merge commit \href{https://github.com/spgroup/mergedataset/blob/master/voldemort/4cc1c145819030c8e2baffe4c92383de14b8d880/source/voldemort/server/VoldemortConfig/merge.java}{de14b8d880}.}
%     \includegraphics[width=\textwidth]{images/fp_harmless_insertion.jpg}
%     \label{fig:fp_harmless_insertion}
% \end{figure*}
%
Harmless code change is similar, but the developers modify existing code instead of inserting new statements.
%
% Esse exemplo de FP é do caso 48
% \begin{figure*}[h]
%     \centering
%     \caption{Harmless code change false positive. The Base version appears on the center. The parents changes appear on the left and the right. Experimental unit from project \emph{Elastic-search-mongodb}, merge commit \href{https://github.com/spgroup/mergedataset/blob/master/elasticsearch-river-mongodb/6b6ce8e851c6613213c4508c3f277a80649e0c7b/source/org/elasticsearch/river/mongodb/Indexer/merge.java}{80649e0c7b}.}
%     \includegraphics[width=\textwidth]{images/fp_harmless_change.jpg}
%     \label{fig:fp_harmless_change}
% \end{figure*}
%
Harmless flow is when there is a valid flow but it existed before integration, that is, it wasn't created by the changes.
Conservative flow is due to the conservative nature of static analysis; it detects a flow that in practice cannot occur, for instance, because our analyses are not object sensitive.
Lastly, there could be annotated related false positives when line information is incorrectly carried on to the bytecode and internal representation.

\begin{figure*}[h]
    \centering 
    \caption{Removed lines false negative. The Base version appears on the center. The parents changes appear on the left and the right. Experimental unit from project \emph{Storm}, merge commit \href{https://github.com/apache/storm/commit/ad2be678831b3b060229fd936e3908110162b7ac}{8110162b7ac}}.
    \includegraphics[width=\textwidth]{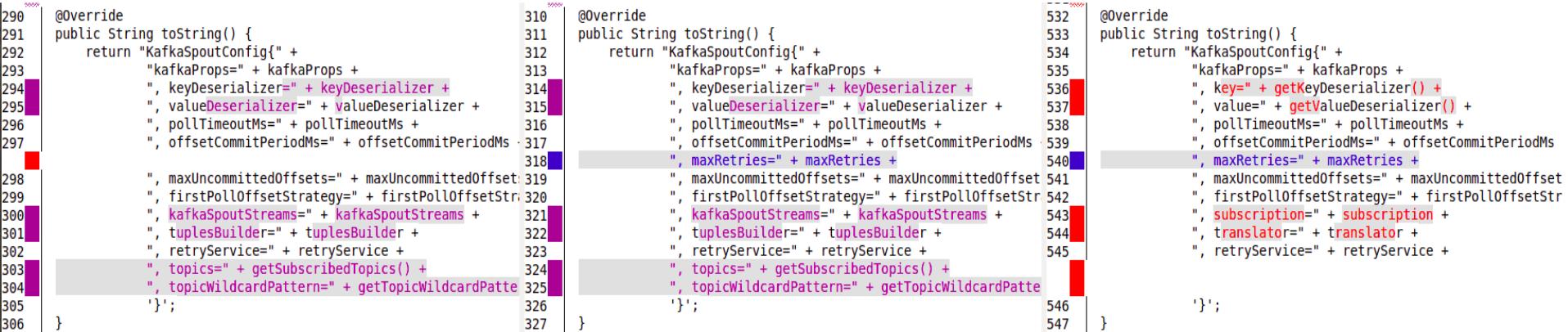}
    \label{fig:fn_example}
\end{figure*}

Moving now to false negatives, one significant reason for misclassification is when the changes made by at least one of the developers consist of line deletions, instead of modifications or insertions.
As our analyses are executed on the Merge commit version, they see no trace of the deleted lines, and therefore cannot compute new flows or overridings caused by the deletions.
Figure~\ref{fig:fn_example} illustrates a case, where both developers delete lines that interfere as they have an effect on the final returned value.  
Our analyses, however, fail to detect interference as no instruction is annotated in this case.
% Utilizamos para este exemplo o cenário 11
This reveals a drawback of our static analysis technique, which relies on the merged version and instruction annotation information associated with the integrated changes (deletions excluded).

A number of false negatives are due to limitations of our analyses current implementation or how we invoke them.
Interface implementation, for example, occurs because the analyses don't exploit global information that could provide potential implementations for an interface or abstract class; so the analyses end up not visiting methods that could reveal the interference. 
% Utilizamos para este exemplo o cenário 76
% \begin{figure*}[h]
%     \centering 
%     \caption{Interface implementation false negative. The Base version appears on the center. The parents changes appear on the left and the right. Experimental unit from project \emph{Elasticsearch}, merge commit \href{https://github.com/spgroup/mergedataset/blob/master/elasticsearch/d896886973660785aac45275ddb110c1a6babc57/source/org/elasticsearch/action/DocWriteResponse/merge.java}{c1a6babc57}.}
%     \includegraphics[width=\textwidth]{images/fn_interface_implementation.jpg} 
%     \label{fig:fn_interface_implementation}
% \end{figure*}
Class dependencies false negatives occur because we run the analysis with JARs that don't include external dependencies, but interference could occur in classes from these dependencies.
% Utilizamos para este exemplo o cenário 51
% \begin{figure*}[h]
%     \centering 
%     \caption{Dependencies false negative. The Base version appears on the center. The parents changes appear on the left and the right. Experimental unit from project \emph{elasticsearch-river-mongodb}, merge commit \href{https://github.com/spgroup/mergedataset/blob/master/elasticsearch-river-mongodb/6b6ce8e851c6613213c4508c3f277a80649e0c7b/source/org/elasticsearch/river/mongodb/Slurper/merge.java}{80649e0c7b}. }
%     \includegraphics[width=\textwidth]{images/fn_dependencies.jpg} 
%     \label{fig:fn_dependencies}
% \end{figure*}
Exception and Recursive methods false negatives respectively result from our analyses not computing exceptional control flows, and analyzing recursive methods just once.
% Utilizamos para este exemplo o cenário 72
% \begin{figure*}[h]
%     \centering 
%     \caption{Exception false negative. The Base version appears on the center. The parents changes appear on the left and the right. Experimental unit from project \emph{Elasticsearch}, merge commit \href{https://github.com/spgroup/mergedataset/blob/master/elasticsearch/d896886973660785aac45275ddb110c1a6babc57/source/org/elasticsearch/action/support/replication/TransportReplicationAction/merge.java}{c1a6babc57}. }
%     \includegraphics[width=\textwidth]{images/fn_exception.jpg} 
%     \label{fig:fn_exception}
% \end{figure*}
% % Utilizamos para este exemplo o cenário 14
% \begin{figure*}[h]
%     \centering 
%     \caption{Recursive method false negative. The Base version appears on the center. The parents changes appear on the left and the right. Experimental unit from project \emph{Jsoup}, merge commit \href{https://github.com/spgroup/mergedataset/blob/master/jsoup/a8b6982de98ff76ef254031d7152fff57f6bf941/source/org/jsoup/helper/HttpConnection/merge.java}{57f6bf941}.}
%     \includegraphics[width=\textwidth]{images/fn_recursive_method.jpg} 
%     \label{fig:fn_recursive_method}
% \end{figure*}
%
Finally, Complex conditions false negatives are caused by a Soot current limitation that doesn't correctly carry on correct line information for conditions involving more than two subexpressions.
% Utilizamos para este exemplo o cenário 65
% \begin{figure*}[h]
%     \centering 
%     \caption{Conditional change false negative. The Base version appears on the center. The parents changes appear on the left and the right. Experimental unit from project \emph{MPAndroidChart}, merge commit \href{https://github.com/spgroup/mergedataset/blob/master/MPAndroidChart/9531ba69895cd64fce48038ffd8df2543eeea1d2/source/com/github/mikephil/charting/renderer/LineChartRenderer/merge.java}{43eeea1d2}. }
%     \includegraphics[width=\textwidth]{images/fn_conditional_change.jpg} 
%     \label{fig:fn_conditional_change}
% \end{figure*}
% \pb{DONE Verificar se cada uma das explicações que dei acima estão exatamente precisas. Essa última mesmo eu sei que precisa de ajuste. Qual ajuste exatamente? RESP.: Está ok, enviei explicação pelo slack. Em resumo:
% O jimple não está pegando corretamente a linha, analisando o bytecode
%    10: if-icmpne     20
%    13: iload-1
%    14: iconst-2
%    15: if-icmpeq     20
%    18: iconst1
% ...
%    LineNumberTable:
% ...
%    line 10: 4
%    line 11: 18
% A linha 11 começa na linha 18 somente, tudo antes disso deveria ser linha 10.

% O Soot está retornando que cada condição está em uma linha diferente: 10, 16 e 13, e deveria ser todas 10.

% if x != 0 goto (branch), 10
% if y != 1 goto (branch), 16
% if y == 2 goto (branch), 13
% x = 1, 11

% Link da issue que abri para o soot: https://github.com/soot-oss/soot/issues/1763
% }
\shortversion{\pb{Galileu, para cada um dos exemplos de FN que aparecem no .tex como comentário, adicionar lá como comentário uma explicação sobre o exemplo em particular, explicando porque nele ocorre FN. POR EXEMPLO, no de dependencies, a falta de que dependência gera o problema, e onde essa dependência aparece na figura. Porque mostrar só a figura não é suficiente para o cara entender.}}{}

\subsection{Performance Results}

\begin{figure}[h]
    \centering
    \caption{Execution time results. Each red dot depicts the average time of 10 executions of our four analyses with one experimental unit. Median is depicted by the red line at 17.8s. For legibility, we omit one outlier, which takes 1421.44 to execute.}
        \label{fig:rain_cloud}
    \includegraphics[width=8.5cm]{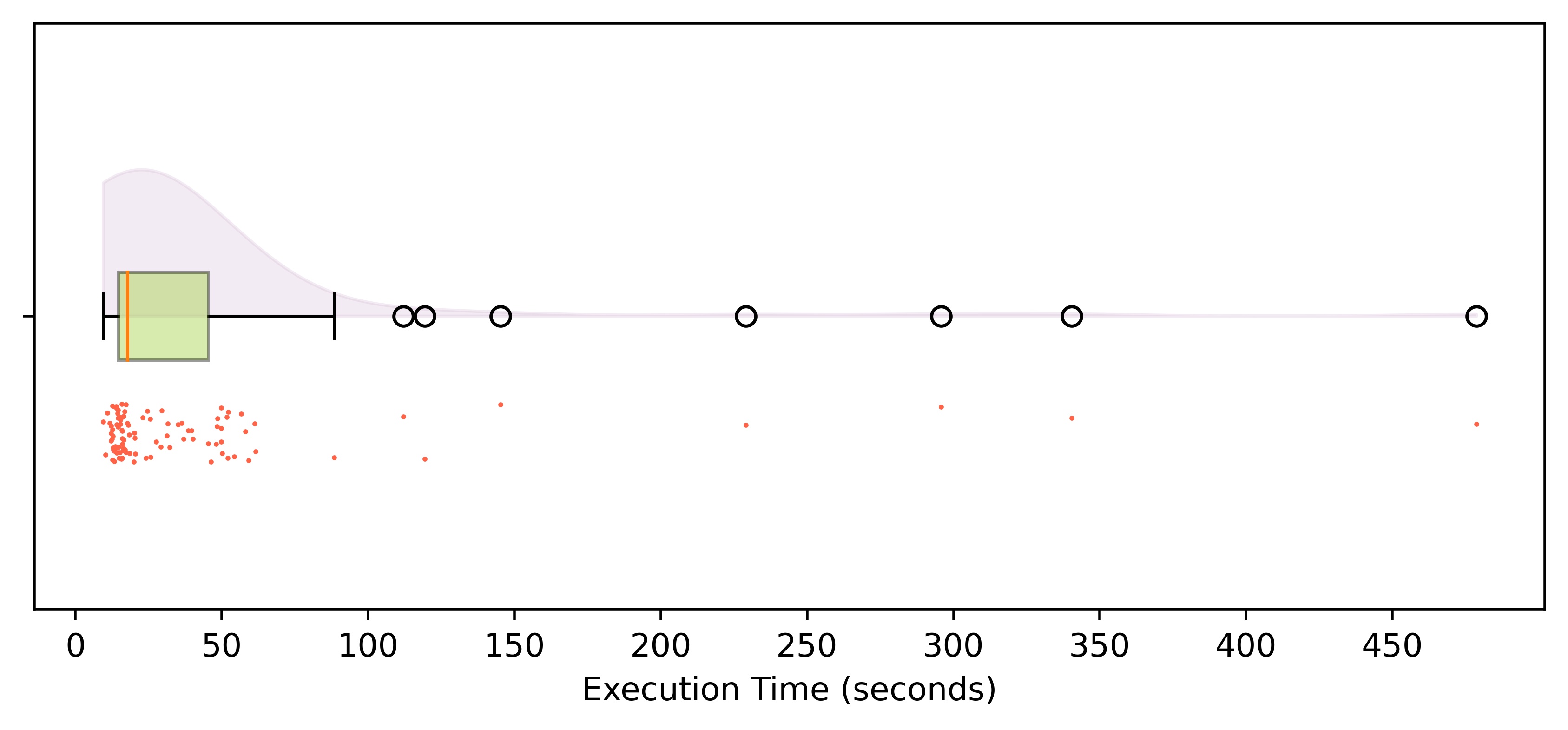}
    %  \pb{DONE Qual o valor exato mesmo? RESP.: A média nesse caso foi 1421.44, mediana: 1417.4, desvio padrão: 21.55}
\end{figure}

We carry out our experiments using the procedures we explained in Section~\ref{sec:method} and 
an Ubuntu docker container running on an Intel Xeon Gold 6338 server with 2Tb of RAM.
Figure~\ref{fig:rain_cloud} shows the average execution time results for each experimental unit (red dots in the figure), as we run our four analysis 10 times for each unit.
Variation across the 10 executions is small, with standard deviation ranging from 0.04s (in a scenario with 11.74s average) to 11.74s (in a scenario with 340.45s average).

The median is at 17.8s for executing together the four analyses, including their setup, graph or abstraction construction, interference checking, and result reporting.
Most units are analyzed in less than 100s, with an implementation that isn't optimized, and could be significantly improved in a number of ways, like by computing a lazy logical disjunction of the analyses, by sharing common setup tasks among the analyses, and by parallelizing analysis execution, as they are independent from each other.
Closely related work benchmarked in the accuracy results either report no performance results, or report much larger execution times.
One can take hours running IF--- which relies on the construction of  System Dependence Graphs~\cite{graf2013using, snelting2015checking, graf2015checking}--- for a single merge scenario; or five minutes only to generate tests with DA, the time to execute them being considerable depending on the project. 
%\pb{DONE Conferir. RESP.: For each tool, we use a budget of 5 minutes and their default configuration. 13 We decide to adopt 5 minutes considering related work opt for different budget configurations (1, 2, or even 10 minutes).}

The depicted outliers, and especially the one that we omitted and that takes a little more than 1400s to execute, consume most time running the interprocedural Data Flow (DF) analysis, both directly and indirectly through Confluence Flow (CF).
%\pb{DONE Quantos \% DF consome?RESP.: DF-right-left consome 186,040s, DF-left-right 185,922s. Confluence SOURCE-BASE 690,968s; SINK-BASE 292,941s; OA: 31,223; PDG-lef: 13,376, right: 11,824
%DF = 26,34\%
%Cf = 69,67%
%CF + DF = 1355,871s, representando 96,00\%
%}
%
They all involve a large number of methods that are visited by DF and CF.
%\pb{DONE Confere? RESP.: Sim. 
%CF visita 1153 métodos
%DF-left visita 581 (22216 nós) e right 572 (21866 nós)
%E o grafo é muito grande, mais de 22 mil nós}

For DF, the number of visits is not directly associated with the size of the entry-point method, although the 1400s outlier has a large entry-point with 235 LOC.
What matters is the number of annotated instructions, which corresponds to the number of lines changed by the developers and integrated in the merge commit.
The more annotated instructions, the more methods are visited and the bigger is the resulting graph.
However, as annotated method calls with DF lead to annotated method bodies, the size of the methods called by the entry-point, and recursively, matters as well.
In the longest outlier, the entry-point method has an annotated method call for a method with 219 LOC, all of which are annotated as well, recursively leading to 570 methods being visited by our analysis.
The deepest level reached in such a case was four, indicating that time was spent in a breadth rather than depth exploration from the entry-point.

For CF, the size of the entry-point method (and recursively the methods it calls) matter, as this analysis essentially visits all called methods.
So the size of the entry-point method in this outlier explains much of the long execution time.

Running the experiment with a depth limit of 10, instead of 5, increases execution time by a factor of 3 to 6, depending of the analysis.
This, however, brings no significant change on accuracy results. 
So we found no need to experiment with other depth limits.

\shortversion{\pb{Eu acredito que o esquema de marcação \emph{eager} de DF pode dar problema se um mesmo método é chamado por left e por Right (Left();...Right();), mesmo que esse método apenas escreva em uma variável e leia outra (x=y). Discutir na reunião com Rodrigo. Trabalho futuro.}}{}

\subsection{Threats to Validity}

\shortversion{Our results validity are subject to the following threats.}{}

\subsubsection*{Construct validity}

Checking for dynamic semantic conflicts is hard with the type of experiment we describe here, as with such design we don't have access to developers desires or even specification they worked against. 
So we follow previous work on dynamic semantic conflicts and adopt a study design that checks for interference, and uses interference ground truth.
There is still, though, a construct threat that is common to previous work: we focus on \emph{locally observable interference}, that is, we consider only interference that can be detected in the method being analyzed and, recursively, in the methods called from it.
It would be hard to manually carry on a global interference analysis considering the whole project.

%Determining the true intention of each developer poses a significant challenge for computational modeling, often making it impractical to use automated tools in this process ~\cite{silva2022detecting}. For this reason, our dataset and ground truth adopt the concept of \emph{locally observable interference} as an approximate technique to identify semantic conflicts.  Also, it is important to mention that our manual analyses do not encompass other forms of globally observable interference, which implies that there may be cases where these interferences exist globally but not locally, or vice-versa.

\subsubsection*{Internal validity}

%When the Left or Right branches remove lines of code, we lose the ability to analyze the merging scenario and perform the necessary analyses, as some vital information may be lost. 

Line annotation as used by our technique is sensitive to the Java compiler version. 
Different compilers might carry on to bytecode slightly different line information, consequently leading to different analyses results.   
We used the JDK Java compiler, as it is likely one of the most widely used.
So our results would hopefully be observed by a significantly number of developers.
%Furthermore, it is essential to highlight that differences in the Java compilers might lead to misleading information that assigns the original line of code at the source code level to the corresponding statements at the Java bytecode level. For example, in our dataset, we did not use the Eclipse compiler, but when conducting tests, we observed significant differences in the accuracy for this assignment. 

As described earlier, some of the scenarios and experimental units in our dataset required manual modifications, especially in configuration files, to compile or to make line information more accurate.
The compilation changes mostly involve dealing with newer dependencies versions; as these are used only to successfully compile the project, and are not included in the analyzed JAR files, there is limited risk of affecting analysis results.   
The line information change in three cases might impact the results; so we say that our experiment relies on three realistic and 96 experimental units.

%These changes were necessary due to imprecisions in mapping lines from the source code derived from bytecode,
%which may not recognize parameters of non-primitive types that have not been previously instantiated in the code.
%The adaptations were manually performed on three scenarios, aiming to preserve the original
%characteristics as much as possible through refactoring, without altering the semantics. 

%Another relevant point is that conditionals with more than two boolean expressions can lead to incorrect assignments of the source line of code in the bytecode statements to the Jimple instructions. Correctly mapping the lines of code to the corresponding Jimple instructions is essential when using the technique we explore in this paper. Therefore, it is necessary to consider these scenarios when interpreting the results, as they can lead to false negatives and directly impact our conclusions.

%Another important point to highlight is that in some scenarios, it was not possible to directly perform the build
%process, so we had to make some modifications to the source code and certain versions of dependencies. However, it is worth noting that no direct changes were made to the merge scenario modifications; they were made in different classes and files,
%thus preserving the scenario in its original form.

\subsubsection*{External validity} Although larger than the datasets used in previous work on dynamic semantic conflicts, our dataset is still limited.
So we can't generalize the obtained results for other samples of merge scenarios and projects, not to mention programming languages. 
Even with a larger dataset, we would still have limited generalization power, as illustrated by the varying results we obtain with the different subsamples used for comparing our technique with each closely related work.

 \section{Discussion}

Based on the accuracy and performance results reported in the previous section, we discuss whether our static analyses could be useful as the basis of semantic merge conflict detection tools.

A tool that first applies textual~\cite{khanna2007formal} or structured merge~\cite{apel2012structured,cavalcanti2017evaluating,seibt2021leveraging} and then runs our analyses on the merged code could benefit from the interference detection capability of the analyses.
This capability is evidenced by the experiment recall results; SA being much superior than DA. 
We expect that such a tool would be more effective than current practices for semantic conflict detection, which are based on code review and running project tests. 
Nevertheless, given our analyses are unable to detect interference in a non neglectable percentage of cases, the tool should be used in addition to the practices, especially when avoiding semantic conflicts is critical.

The precision results, though, show that these benefits would come with potentially significant costs for handling false positives from an SA based tool.
The costs would be inferior than if using IF, for example, and could nevertheless be worth in more critical contexts where letting conflicts escape to the production environment can be harmful or costly.
In less critical contexts, a DA based tool--- with better precision, but much worse recall --- could be more adequate, in spite of the higher computational costs of running test based dynamic analysis.

The false positives costs could be likely reduced by a tool that is also able to detect refactoring changes~\cite{silva2017refdiff, tsantalis2020refactoringminer2}.
If all the changes from one developer are structural refactorings, the tool can simply report no interference; no need for calling the analyses in this case.
Otherwise, the tool can invoke the analyses and filter out resulting flows if either the start or the end of the flow is an instruction associated with a refactoring change.
This could avoid a significant part of the false positives in our sample.
We need, however, further study to understand how the inaccuracies of SA interplay with those from refactoring detection tools.
%no simple bytecode equivalence tool would be useful.
%Falar que comparamos todos os .jar e .class dos cenários que são refactoring e não detectamos nenhum arquivo igual no

Making the analyses more accurate could also help to reduce the false positive costs, by reducing the conservative flows we report in the previous sections.
But we expect no major impact from that. 
The other false positive categories (harmless code insertion, change, and harmless flow, etc.) are inherent to our technique and cannot be avoided.

Most false negatives we observed are due to limitations of our analyses current implementation or how they are invoked.
Improving that could increase the potential of an SA based tool.
But there is a complex trade-off between such improvements and their impact on performance and false positives, and further study is necessary to understand that.
%recursive methods, exception, interface implementation, class dependencies could be solved by improving the analyses or how they are invoked. at the expense of performance.
The removed lines false negatives are inherent to our technique, and the complex condition one can be eliminated by fixing the Soot analysis infrastructure we use.
% depends on new technique for computing graphs and taking differences.

The performance results suggest that an SA based tool has the potential to be invoked for most merge situations, especially considering that the current analyses implementation has plenty of room for optimization, as discussed in the previous section.
Although analysis depth is controlled by our implementation, breadth is not and could be a problem for a few cases, as the discussed outliers show.
In these cases, background checking as, for instance, with continuous integration servers would be an option.
A SA based tool, though, should have much better performance than an IF, DA, and likely a VA tool as well.

A semantic merge conflict detection tool could not be restricted to changes in the same operation, as considered in our experiment. 
Handling such situations would simply require an algorithm for finding or creating a proper entry point that captures the integrated changes.
%most close common calling place.
This, however, could require running the analyses with different depth levels, leading to longer execution times.

Although we have focused on Java projects, a large part of our analysis infrastructure could be applied to other languages compiled to Java bytecode.
Most ideas could also be applied to other languages.
Results would be likely inferior for dynamic languages.
For such languages, DA could be a more promising alternative.
This technique can also capture interference due to changes in field declarations with initializations~\cite{da2020detecting}.
To capture that, a SA based tool would require a simple preprocessing transformation for creating an entry-point method containing the initialization code.
Similarly, extract variable transformations would help to to avoid the line annotation issues we mentioned in Section~\ref{sec:method}.

% code review tool could benefit

 \section{Related Work}

%% Silva and Borba investigate to better understand build conflicts \cite{da2022build}. Revealed by failures when integrating code, they investigated their frequency, structure, and resolution patterns in 451 open-source Java projects from GitHub, analyzing the Travis build logs and checking if error messages are related to merged changes. They found and classified 239 build conflicts and their resolution patterns. On the other hand, Dias and Borba, investigate seven critical factors related to modularity, size, and developers' contribution time in the context of preventing software conflicts\cite{dias2020understanding}. With a dataset of 73504 merge scenarios from GitHub, encompassing projects developed in Ruby and Python, the obtained results indicate that the probability of merge conflicts significantly increases when the contributions to be integrated lack modularity and are made by multiple developers with numerous commits and altered files. Some more recent studies \cite{de2003breaking, ghiotto2018nature} have shown that the merging process can generate bugs in the code and delay the software development cycle. In our technique, we do not analyze the modularity of scenarios but rather changes made within the same method, with the same objective of preventing conflicts. These works focus on textual conflicts and present their reasons and solutions. In our case, we use static analysis to try to identify scenarios that have semantic conflicts.

% \subsection*{Dynamic Semantics} 

Horwitz et al. first studied dynamic semantic conflicts~\cite{horwitz1989integrating,horwitz1990interprocedural,yang1992program} and formalize the interference definition we use here.
They, however, propose the construction of PDGs~\cite{horwitz1989integrating} and SDGs~\cite{horwitz1990interprocedural} for the four program versions in a merge scenario, and compute differences between these to detect and resolve interference; to compute differences, they rely on the use structured editors to identify AST nodes changed in each of the revisions.
Although inspired by this seminal work, and also using PDGs, our technique only analyzes the merge version--- which is annotated with line changing information--- and avoids some of the heavyweight analysis needed to construct SGDs. 
By not dealing with four graphs as they do, we can't detect the effects that line deletions might have on interference.
On the other hand, we gain in efficiency, and likely in precision, but losing in recall.
We can't know for sure as they don't empirically evaluate their technique, nor provide implementations.
More recent work~\cite{barros2017using} gets close to that, using the JOANA framework~\cite{joana-paper} to build a single SDG for the merge version in a scenario.
This, however, can take hours to construct, whereas our analyses run much faster.
In Section~\ref{sec:results}, we compare this in more detail with our approach, showing experimental results.
Overall, our technique presents better but comparable precision; we were not able to compare recall due to the related work experiment design. 

Instead of static analysis, Silva et al.~\cite{da2020detecting} detects interference by using a dynamic analysis based on automated unit test generation.  
They rely on test passing criteria that suggests interference--- if a generated test breaks with the Base version, and passes in one of the parents, and breaks again in the merge is evidence that the change carried on by this parent is not preserved in the merge version.
They need no line change information, and can detect interference due to removed lines and changes in field declarations.
We compare this in more detail in Section~\ref{sec:results}. 
Overall, our technique presents much better recall but much inferior precision.
Our performance should also be much better than theirs, based on the little information (only test generation time, not test execution time) they provide about that.
The combination of both techniques could be a promising direction for future work.

A third technique for interference detection is explored by Sousa et al.~\cite{sousa2018verified}, which statically infers relational post-conditions from the merge scenario code versions, establishing constraints on how state elements can be modified by different versions in such a way to avoid interference.
It then applies theorem proving, as in SMT solving, to check if the constraints are satisfiable.
We again compare this in more detail in Section~\ref{sec:results}, showing comparable precision to our approach; we were not able to compare recall due to the related work experiment design. 
Performance from this verification approach can be considerably higher than our approach with larger programs that were changed in a number of places.

More recently, researchers have applied large language models to detect integration conflicts~\cite{zhang2022using}. 
Although their motivation involves the three kinds of conflicts we discuss in the introduction, in the paper they report results only about textual and static semantic (build) conflicts.
They claim that dealing with dynamic semantic conflicts would be hard, especially for establishing ground truth to train and test such models.
With static semantic conflicts, they obtain slightly higher precision than we obtain here with dynamic semantic conflicts.
%
%For instance, Zhang and Todd investigated the feasibility of using pre-trained language models to identify semantic and textual merge conflicts  \gs{Ele utiliza conflito de build para isso, então não dynamic analysis}. The authors implemented the Gmerge tool, which automatically suggests merging conflict resolutions. It takes a merge conflict and merge histories from both upstream and downstream as input and returns a conflict resol%ution plan that indicates which lines of code need to change. The authors evaluated their technique using Microsoft Edge and achieved an accuracy of 64.6\%---which suggests their technique is promising. 
%
%\gs{Aqui, dá pra falar que o LOI de zhang2022using foi bem diferente do nosso: In one of the examples, they show a case where there was a renaming, \code{IsIncognito()} was renamed to \code{GetIncognito()}, categorizing it as a semantic conflict. However, in our ground truth, this type of change is not considered interference, as it is a refactoring and does not alter the system's behavior.}
%
Brun et al.~\cite{brun2013early} also discuss semantic conflicts, but mostly collect evidence of static semantic conflicts. 
They use project tests to detect dynamic semantic conflicts.
Previous work~\cite{sarma2011palantir,apel2011semistructured,apel2012structured,cavalcanti2017evaluating,cavalcanti2019impact,da2022build} tries to prevent, detect, or resolve textual and static semantics conflicts.

 \section{Conclusions}

\shortversion{\pb{Ler com cuidado Preparando as referências bibliográficas em artigo na minha página. Ver link do artigo no .tex. E ajustar o .bib de acordo. No momento não está sendo seguido um padrão.} https://pauloborba.cin.ufpe.br/post/escrevendo-seu-trabalho/}{}

We present a technique and a set of static analyses that show significant interference detection capability, and potential for detecting dynamic semantic conflicts.
It comes, though, with potentially significant costs for handling false positives.
Comparing with previous work~\cite{barros2017using, sousa2018verified, silva2022detecting}, our technique shows much better F1 score and recall than the dynamic analysis technique, but with much worse precision.
Our precision is comparable to the ones presented by the SDG and theorem proving techniques.
We also manually analyze and classify our analyses' false positives and negatives, and discuss how, for instance, our precision could be significantly improved if we applied refactoring detection techniques before running the analyses. 
The performance results show that our analyses should have much better performance than previous techniques.
In summary, besides existing limitations, our experience suggests that our technique is feasible, and might complement dynamic analysis alternatives that aim to detect interference.

\shortversion{ and help to identify semantic conflicts}{}

\begin{acks}
For partially supporting this work, we would like to thank INES (National Software Engineering Institute) and the Brazilian research funding agencies CNPq, FACEPE (grants IBPG-0029-1.03/20), and CAPES.
SPG, Christian, colegas que ajudaram com dataset, mining, análises, ferramentas, etc.
\end{acks}

\bibliographystyle{ACM-Reference-Format}
\bibliography{sample-base}

\end{document}